\documentclass[]{elsarticle}

\usepackage{hyperref}

\journal{Journal of Parallel and Distributed Computing}

\usepackage[toc,acronym]{glossaries}
\usepackage{xcolor}
    \definecolor{webyellow}{rgb}{0.98,0.92,0.73}
\usepackage{booktabs}
\usepackage{ifthen}
\usepackage{listings}
\usepackage{tikz}
\usetikzlibrary{arrows,shadows,backgrounds,calc,decorations.pathreplacing,decorations.pathmorphing,positioning, 
	automata}
	\usepackage{pgfplots}
	\pgfplotsset{compat=1.13} 
	\usepackage{pgfplotstable}
	\usepackage{adjustbox}
\definecolor{darkmagenta}{rgb}{0.55, 0.0, 0.55}
\usepackage{xkeyval}
\usepackage{tikz}
\usetikzlibrary{
	shapes,
	positioning,
	arrows,
	fit,
	backgrounds,
	calc,
	tikzmark,
	shadows,
	automata}
\makeglossaries

  \newacronym{ASIC}{ASIC}{Application Specific Integrated Circuit}
  \newacronym{CMC}{CMC}{Configurable Architecture \gls{MC}}
  \newacronym{CPU}{CPU}{Central Processing Unit}
  \newacronym{EMPA}{EMPA}{Explicitly Many-Processor Approach}
  \newacronym{FPGA}{FPGA}{Field Programmable Gate Array}  
  \newacronym{HW}{HW}{hardware}
  \newacronym{HPL}{HPL}{High Performance Linpack}
  \newacronym{HPCG}{HPCG}{High Performance Conjugate Gradients}
  \newacronym{ISA}{ISA}{Instruction Set Architecture}
  \newacronym{I/O}{I/O}{Input/Output}
  \newacronym{MC}{MC}{Multi-Core and/or Many-Core}
  \newacronym{MLP}{MLP}{Memory Level Parallelism}
  \newacronym{OoO}{OoO}{Out-of-Order}
  \newacronym{OS}{OS}{operating system}
  \newacronym{PD}{PD}{Propagation Delay}
  \newacronym{PU}{PU}{Processing Unit}
  \newacronym{QT}{QT}{quasi-thread}
  \newacronym{PC}{PC}{Program Counter}
  \newacronym{RC}{RC}{reconfigurable}
  \newacronym{RT}{RT}{real-time}  
  \newacronym{SPA}{SPA}{Single Processor Approach}
  \newacronym{RMC}{RMC}{Reconfigurable Architecture \gls{MC}}
  \newacronym{SV}{SV}{supervisor}
  \newacronym{SW}{SW}{software}

	\makeatletter
	\define@key{MEMacros}{color}[green]{\def\ME@color{#1}}
\makeatother
\usepackage{etoolbox}

\makeatletter
\newcommand{\highlighted}[2][]{%
	\setkeys{MEMacros}{color=darkmagenta!40!black!80,#1}%
	{\textit{\textbf
			{\textcolor{\expandafter\ME@color\expandafter}{#2}}}}%
}
\makeatother 

\makeatletter
\newcommand{\gettikzxy}[3]{%
  \tikz@scan@one@point\pgfutil@firstofone#1\relax
  \edef#2{\the\pgf@x}%
  \edef#3{\the\pgf@y}%
}
\makeatother

\pgfplotscreateplotcyclelist{my color list}{%
solid, color=webblue, every mark/.append style={solid, fill=webblue}, mark=*\\%
densely dashdotted, color=webred, every mark/.append style={solid, fill=webred},mark=diamond*\\%
densely dotted, color=webgreen, every mark/.append style={solid, fill=webgreen}, mark=triangle*\\%
loosely dashed, color=webbrown, every mark/.append style={solid, fill=webbrown},mark=square,fill=webgreen!20\\%
densely dotted, every mark/.append style={solid, fill=gray}, mark=otimes*\\%
dashed, every mark/.append style={solid, fill=gray},mark=diamond*\\%
densely dashed, every mark/.append style={solid, fill=gray},mark=square*\\%
dashdotted, every mark/.append style={solid, fill=gray},mark=otimes*\\%
dashdotdotted, every mark/.append style={solid},mark=star\\%
}
	\definecolor{webgreen}{rgb}{0,.5,0}
	\definecolor{webbrown}{rgb}{.6,0,0}
	\definecolor{webyellow}{rgb}{0.98,0.92,0.73}
	\definecolor{webgray}{rgb}{.753,.753,.753}
	\definecolor{webblue}{rgb}{0,0,.8}
    \definecolor{webgreen}{rgb}{0, 0.5, 0} 
    \definecolor{webred}{rgb}{0.5, 0, 0}   

\tikzstyle{QT} = [top color=white, bottom color=blue!30,
minimum height={0.8},rotate=90,scale=\scalefact,xshift=-.5cm,yshift=-.5cm,
draw=blue!50!black!100, drop shadow] 

\tikzstyle{QTC} = [top color=green!80, bottom color=white,
minimum height={0.8},scale=\scalefact,xshift=-.5cm,yshift=-.8cm,
draw=blue!50!black!100, drop shadow] 
\newcommand\QTfigure[6]   
{
	\draw
	($ (0 cm,0.5cm-.5*#2 cm) + (0 cm,-#5*.1 cm) +(#3cm,-#4cm)$) node[QT,minimum width=#2cm,draw]
	(Q#1) {$#1$}
	;
} 
\newcommand\QTCfigure[6]   
{
	\draw
	($ (3.0 cm,.6cm-.5*#2 cm) + (0 cm,-#5*.1 cm) +(#3cm,-#4cm)$) node[QTC,minimum width=5cm,minimum height=#2cm,draw]
	(Q#1) {$#1$}
	;
} 

\newcommand\QTAfigure[6]   
{
	\draw
	($ (2.7cm 
	,
	.6cm-.5*#2 cm) + (0 cm,-#5*.1 cm) +(#3cm,-#4cm)$) node[QTC,minimum width=3cm,minimum height=#2cm,draw]
	(Q#1) {${#1}$}
	;
} 

\newcommand\QTRfigure[6]   
{
	\draw
	($ (0 cm,0.5cm-.5*#2 cm) + (0 cm,-#5*.1 cm) +(#3cm,-#4cm)$) node[QT,color=red,minimum width=#2cm,draw]
	(Q#1) {$#1$}
	;
} 
\def\scalefact{1}

\begin{document}

\begin{frontmatter}

\title{Limitations of performance of Exascale Applications\\
and supercomputers they are running on}

\author{{J\'ANOS V\'EGH}
\address{University of Miskolc, Hungary}}

\begin{abstract}
The paper highlights that the cooperation of the components
of the computing systems receives even more focus in the
coming age of exascale computing.
It discovers that  inherent performance limitations exist
and identifies the major critical contributions of the
performance on many-many processor systems.
The extended and reinterpreted simple Amdahl model describes 
the behavior of the existing supercomputers surprisingly well, and explains some mystical happenings around high-performance computing. 
It is pointed out that using the present technology
and paradigm only marginal development of performance is possible, and that the major obstacle towards higher performance applications is the 70-years old computing paradigm itself. A way to step forward is also suggested.
\end{abstract}


\end{frontmatter}



\section{Introduction}

In the age of exascale applications new challenges
for the application writers and users appear. 
The supercomputers --and especially  the exascale ones-- are strongly custom-made, rather than general purpose computers.
The cooperation of the components will be 
as crucial as it was never before, so the 
users and the programmers  of the exascale applications
--as well as the architecture designers--
must exactly understand how the strongly parallelized and distributed computing system works.
Otherwise minor imprecisions, a poorly organized cooperation or a small amount
of poorly organized code can result in 
catastrophics performance losses.

With the expected and wanted advent of exascale computing a lot of 
application developments has been started practically in all fields of science,
military, industry, services, etc. The increased interest is partly motivated
by real economical needs (as proved by the growing number of industry-hosted supercomputers),
by the need of elaborating "big data", by the increased importance of computer modeling
or providing platform for applications utilizing artificial intelligence, etc.
In contrast with those "commodity supercomputers", the "racing supercomputers"
are mainly motivated by the prestige value of the better position on the list of
high-performance supercomputers, "green" supercomputers or total supercomputer capacity.
Of course the acquired experiences and developed technologies can be effectively 
utilized later in the "commodity supercomputers" as well.
The ongoing race between institutions, nations, companies, processors,
interconnection and memory access methods, etc. resulted in a kind of
"gold rush". The computing performance quietly approached its technological bounds,
but the "computing stack" has not been revised, in the past 70 years.

It is well known, that computing has its numerous limitations and even those limitations
are limited~\cite{LimitsOfLimits2014}, as well as that the computing growth
(like Moore's observation) is only initially exponential~\cite{ExponentialLawsComputing:2017}.
It is also a common experience on all fields that when approaching
some extremities (big masses, small sizes, big sizes, high speed), the behavior
of the studied subject drastically changes.
Followers of the high-performance computing field faced strange experiences, like
aborting projects in the very last phase of their development, 
utilizing only a fragment of the available nominal performance
when nominating to the TOP500~\cite{TOP500:2017},
withdrawn from the competition in a half year,
different ranking on different benchmarks,
tragically low efficiency of some application on some architecture,
strategic allience of manufacturers with conflicting interests for
producing new processor/accelerator/supercomputer,
or calling for project ideas for a supercomputer with unknown
architecture and features.
All these happenings seem to support  the presumption that 
parallel processing has approached some mystical performance bound.

Amdahl~\cite{AmdahlSingleProcessor67} called first the attention to that \textit{the parallel systems built from single processors have serious performance limitations}.
His warning was used successfully on quite different fields~\cite{UsesAbusesAmdahl:2001},
sometimes misunderstood, abused and even refutated (for a review see~\cite{AmdahlsRefutation_Devai2017}).
Despite of that 
"\textit{Amdahl's Law is one of the few, fundamental laws of computing}"~\cite{AmdalsLaw-Paul2007}, for today it is quite forgotten
because it is commonly assumed to be valid for \gls{SW} only, 
although Amdahl was speaking about complex computing systems.

The paper attempts to review some issues important for the coming
exascale computing age and is structured as follows. 
Amdahl's law is reinterpreted for modern computing systems in section~\ref{sec:reinterpret} in its original spirit, rather than
as formulated by the successors of Amdahl and commonly accepted today.
Section~\ref{sec:InherentLimit} explains why in consequence of Amdahl's Law the present computing
technologies have an inherent (although depending on many factors) upper bound for the achievable performance gain.
In the light of this, section~\ref{sec:SupercomputerHistory} considers the documented history of supercomputers,
and demonstrates that the efficiency of parallelization (in other words the achievable parallelization performance gain)
governed the supercomputer development.
Based on the idea of Amdahl that the different non-parallelizable contributions,
\textit{independently of their origin}, act as a single non-parallelizable fraction,
section~\ref{sec:ParallelModel} introduces a by intention strongly simplified model,
that assigns some separated fractions to some major contributors, enabling to prepare
a semi-technical model.
The behavior of the introduced contributions is analyzed in section~\ref{sec:Contributions}.
Section~\ref{sec:Benchmarking} explains what attribute of supercomputers is measured by the
different benchmarks and why selecting a proper benchmark is important for ranking the supercomputers.
As section~\ref{sec:Efficiency} presents, the model enables to understand what
factors affect the efficiency of an application on supercomputers having different configurations.
The behavior of applications depend heavily on the supercomputer \gls{HW}.
Based on the behavior of those different contributions, section~\ref{sec:Perspectives}
provides some short-term predictions on the development of supercomputer performance.
As it can be concluded from the previous sections, and was  guessed by many researchers,
one of the major obstacles on the road towards even higher performance is the computing paradigm itself.
This issue is discussed in section~\ref{sec:Paradigm}, where also a possible solution is suggested.
The discussion is concluded in section~\ref{sec:Summary} with repeating the prophecy of Amdahl: 
\textit{the age of single-processor approach development is over, other than marginal advances will only be possible
if using cooperating processors}.

\section{Reinterpreting Amdahl's Law}\label{sec:reinterpret}

\subsection{Amdahl's idea}
A general misconception (introduced by the successors of Amdahl) is to assume that Amdahl's law
is valid for software only and that the non-parallelizable fraction contains
something like the ratio of the numbers of the corresponding instructions.
Actually,\textit{ Amdahl's law is valid for any partly parallelizable activity (including computer unrelated ones) and the non-parallelizable fragment  shall be given as the ratio of the time spent with non-parallelizable activity 
to the total time}.
Amdahl in his famous paper~\cite{AmdahlSingleProcessor67} speaks about
"\textit{the  fraction  of  the  computational  load}"
and explicitly mentions, in the same sentence and same rank, algorithmic reasons like 
"\textit{computations  required   may  be  dependent  on  the  states  of  
the   variables   at   each   point}"; architectural aspects like "\textit{may  be strongly   dependent   on  sweeping  through  the  array  
along  different   axes  on  succeeding  passes}"
as well as
"\textit{physical problems}" like 
"\textit{propagation   rates   of  different  physical  effects may be quite different}".
His point of view is valid also today: 
one has to consider \textit{the workload of the complex \gls{HW}/\gls{SW} system,
rather than some segregated component}, and his idea describes parallelization imperfectness of any kind.
\textit{When applied to a particular case, especially in the case of exascale systems,  it shall be scrutinized which contributions can be neglected. 
Notice that the eligibility of neglecting some component changes with time, technology and conditions.}

\subsection{Deriving the effective parallelization}
Successors of Amdahl expressed Amdahl's law  with the formula 

\begin{equation}
S^{-1}=(1-\alpha) +\alpha/k \label{eq:AmdahlBase}
\end{equation}

\noindent where $k$ is the number of parallelized code fragments, 
$\alpha$ is the ratio of the parallelizable part within the total code,
$S$ is the measurable speedup. 
The assumption can be visualized that (assuming many processors)
in $\alpha$ fraction of the running time the processors are executing parallelized code,
in (1-$\alpha$) fraction they are waiting (all but one), or making non-payload activity.
That is $\alpha$ describes how much, in average, processors are utilized,
or how effective (at the level of the computing system) the parallelization is.

For a system under test, where  $\alpha$ is not \textit{a priory} known,
one can derive from the measurable speedup  $S$ 
an \emph{effective parallelization} factor~(\cite{Vegh:2017:AlphaEff}) as

\begin{equation}
\alpha_{eff} = \frac{k}{k-1}\frac{S-1}{S} \label{equ:alphaeff}
\end{equation}

\noindent Obviously, this is not more than $\alpha$ expressed in terms of $S$ and $k$ from Equ.~(\ref{eq:AmdahlBase}).
For the classical case, $\alpha = \alpha_{eff}$; which simply means that
in the \emph{ideal} case the actually measurable effective parallelization 
achieves the theoretically possible one.
In other words, $\alpha$ describes a system the \emph{architecture} of which is completely known,
 while $\alpha_{eff}$ characterizes the \emph{performance}, which describes both the complex architecture and the actual conditions. 
 It was also demonstrated~\cite{Vegh:2017:AlphaEff} that  $\alpha_{eff}$ 
 can be successfully utilized to describe parallelized behavior from \gls{SW} load balancing
 through measuring efficiacy of the on-chip \gls{HW} communication to characterize
 performance of clouds.

The value $\alpha_{eff}$ can also be used to refer back to Amdahl's classical
assumption even in the realistic case when the parallelized chunks 
have different lengths and the overhead to organize parallelization is not negligible.
The speedup $S$  can be measured and $\alpha_{eff} $ can be utilized
to characterize the measurement setup and conditions, 
how much from the theoretically possible maximum parallelization is realized.
Numerically ($1-\alpha_{eff}$) equals with the $f$ value, established theoretically~\cite{Karp:parallelperformance1990}.

The distinguished constituent in Amdahl's classic analysis is the parallelizable fraction $\alpha$,
all the rest (including wait time, non-payload activity, etc.) goes into the "sequential-only" fraction.
When using several processors, one of them makes the sequential calculation, the others are waiting
(use the same amount of time). So, when calculating the speedup, one calculates

\begin{equation}
S=\frac{(1-\alpha)+\alpha}{(1-\alpha)+\alpha/k} =\frac{k}{k(1-\alpha)+\alpha}
\end{equation}
hence the  efficiency is
\begin{equation}
E = \frac{S}{k}=\frac{1}{k(1-\alpha)+\alpha}\label{eq:soverk}
 \end{equation}

\noindent This explains the behavior of diagram $\frac{S}{k}$ in function of $k$ experienced in practice:
the more processors, the lower efficiency.
At this point one can notice that $\frac{1}{E}$ is a linear function of  number of processors, and its slope equals to $(1-\alpha)$.
Equ.~(\ref{eq:soverk}) also underlines the importance of the single-processor performance:
the lower is the number of the processors used in the parallel system having the expected performance,
the higher can be the efficacy of the system.

Notice also, that through using Equ.~(\ref{eq:soverk}), the efficiency $\frac{S}{k}$ can be equally good for describing the efficiency of parallelization of a setup,
provided that the number of processors is also known. From Equ. (\ref{eq:soverk})
\begin{equation}
\alpha_{E,k} = \frac{E k -1}{E (k-1)}\label{eq:alphafromr}
\end{equation}

\noindent 
If the parallelization is well-organized (load balanced, small overhead, right number of processors), $\alpha_{eff}$  is close to unity,
so tendencies can be better displayed through using $(1-\alpha_{eff})$ in the diagrams below.

The importance of this practical term $\alpha_{eff}$ is underlined by that
the achievable speedup (performance gain) can easily be derived from Equ.~(\ref{eq:AmdahlBase}) as

\begin{equation}
G=\frac{1}{(1-\alpha_{eff})} \label{eq:AmdahlMax}
\end{equation}

\subsection{The original assumptions}
The classic interpretation implies three\footnote{An additional essential point which was missed by both~\cite{Karp:parallelperformance1990} and
~\cite{ExponentialLawsComputing:2017},
that \textit{the same computing model was used in all computers considered}.} 
 essential restrictions,
but those restrictions are rarely mentioned in the textbooks on parallelization:
\begin{itemize}\setlength\itemsep{0em}
	\item the parallelized parts are of equal length in terms of execution time
	\item the housekeeping (controling parallelization, passing parameters,
	waiting for termination, 
	exchanging messages, etc.) has no cost in terms of execution time
	\item the number of parallelizable chunks coincides with the number of available computing resources
\end{itemize}
\noindent Essentially,  this is why \emph{Amdahl's law represents a theoretical upper limit for parallelization gain}.
It is important to notice, however, that\textit{ a 'universal' speedup exists only if the parallelization 
efficiency $\alpha$ is independent from the number of the processors}. As will be discussed in section~\ref{sec:Contributions},
this assumption is only valid if the number of processors is low, so the usual linear extrapolation of the actual performance  
on the nominal performance will not be valid any more in the case of the exascale computing systems.

\subsection{The additional factors considered here}
In the spirit of the
\textit{Single Processor Approach} (SPA) the programmer (the person or the compiler) has to organize the job:
at some point the initiating processor splits the 
execution, transmits the necessary parameters to some other processing units, starts
their processing, then waits for the termination of started processings; see section~\ref{sec:InherentLimit}.
Real-life programs show sequential-parallel behavior, with variable degree
of parallelization~\cite{YavitsMulticoreAmdahl2014} and even apparently 
massively parallel algorithms change their behavior during processing~\cite{Pingali:2011:TaoOfParallelism}.
All these make Amdahl's original model non-applicable, and call for extension.

As discussed in~\cite{LimitsOfLimits2014}
\begin{itemize}
\item many parallel computations today
are limited by several forms of communication and synchronization
\item the parallel and sequential runtime components are only slightly affected
by cache operations
\item  wires get increasingly slower relative to gates
\end{itemize}
\noindent In the followings
\begin{itemize}
\item the the main focus will be on synchronization and communication; they are kept at their strict absolute minimum; and their effect is scrutinized
\item the effect of cache will be neglected, and runtime components not discussed separately
\item the role of the wires is considered in an extended sense: both the importance of physical distance and using special connection methods will be discussed
\end{itemize}

\section{The inherent limit of parallelization}\label{sec:InherentLimit}

As it was mentioned in the previous section,
initially and finally only one thread exists,
i.e. the minimal absolutely necessary non-parallelizable activity is to fork the
other threads and join them again.
With the present technology, no such actions can be shorter than one clock period.
That is, the non-parallelizable fraction will be given as the ratio of the time of the two clock periods to the total execution time.
The latter time is a free parameter in describing the efficiency, i.e. the value of the effective parallelization $\alpha_{eff}$ also depends
on the total benchmarking time (and so does the achievable parallelization gain, too).

This dependence is of course well known for supercomputer scientists: for measuring efficiency with better accuracy (and also for producing better $\alpha_{eff}$ values)
hours of execution times are used in practice. For example in the case of benchmarking the supercomputer
$Taihulight$~\cite{DongarraSunwaySystem:2016}  13,298 seconds benchmark runtime was used; on the 1.45 GHz processors it means 
$2*10^{13}$ clock periods. This means that (at such benchmarking time) the inherent limit of  $(1-\alpha_{eff})$
is $10^{-13}$ (or equivalently the achievable performance gain is $10^{13}$).
In the followings for simplicity 1.00 GHz processors (i.e. 1 ns clock cycle time) will be assumed.

The supercomputers, however, are distributed systems.
In a stadium-sized supercomputer the distance between processors (cable length) about 100 m can be assumed.
The net signal round trip time is cca. 
$10^{-6}$ seconds, or $10^{3}$ clock periods.
The presently available network interfaces have 100\dots200 ns latency times, and sending a message between processors 
takes time in the same order of magnitude.
This also means that \textit{making better interconnection 
is not a bottleneck in enhancing performance}.
This statement is underpinned also by statistical considerations~\cite{Vegh:StatisticalConsiderations:2017}.

Taking the (maybe optimistic) value $2*10^{3}$
clock periods for the signal propagation time,
the  value of the effective parallelization   $(1-\alpha_{eff})$ will be at least in the range of $10^{-10}$,
only because of the physical size of the supercomputer. 
This also means that the expectations against the absolute performance of supercomputers are excessive: assuming a 
10 Gflop/s processor, the achievable absolute \textit{nominal} performance is $10^{10}$*$10^{10}$, i.e. 100 EFlops.
Because of this, in the name of the company PEZY
\footnote{https://en.wikipedia.org/wiki/PEZY\_Computing: The name PEZY is an acronym derived from the greek derived Metric prefixs Peta, Eta, Zetta, Yotta
}
the last two letters are surely obsolete.
It looks like that in the feasibility studies  an analyzis
for whether this inherent performance bound exists is done neither in USA~\cite{ScienceExascaleRace:2010} nor EU\cite{EUActionPlan:2016}.

Another major issue arises from the computing principle 
\gls{SPA}: only one computer at a time can be addressed
by the first one. As a consequence, minimum as many 
clock cycles are to be used for organizing the parallel work as many addressing steps required.
Basically, this number equals to the number of cores 
in the supercomputer, i.e. the addressing in the TOP10 positions typically needs clock cycles in the order of
$5*10^{5}$\dots$10^{7}$;
degrading the value of $(1-\alpha_{eff})$ into the range
 $10^{-6}$\dots$2*10^{-5}$.
Two tricks may be used to reduce the number of the addressing steps:
either the cores are organized into \textit{cluster}s
as many supercomputer builders do, or the processor itself can
take over the responsibility of addressing its cores~\cite{CooperativeComputing2015}.
Depending on the actual construction, the reducing factor
can be in the range $10^{2}$\dots$5*10^{4}$,
i.e the resulting value of $\alpha_{eff}$ is expected to be
in the range of $10^{-8}$\dots$2*10^{-6}$.
Notice that utilizing "cooperative computing"~\cite{CooperativeComputing2015} enhances further
the value of $(1-\alpha_{eff})$, but it means already utilizing a (slightly) different computing principle.

An operating system must also be used, for protection and convenience. If one considers the context change with its consumed $2*10^4$ cycles~\cite{Tsafrir:2007},
the absolute limit is cca. $5*10^{-8}$, on a zero-sized supercomputer. This value is also very close to the "danger zone" derived above.
This is why $Taihulight$ runs the actual computations in kernel mode~\cite{CooperativeComputing2015}.

\textit{It is crucial to understand that the decreasing efficiency (see Equ.~(\ref{eq:soverk})) is coming
from the computing principle itself rather than from some kind of engineering imperfectness.
This inherent limitation is of principial nature and cannot be mitigated without changing the computing principle.}
For validating these limitations, see also the measured performance data in Fig.~\ref{fig:exaRMaxAlpha}.

Although not explicitly dealt with here, 
notice that the data exchange between the first thread
and the other ones also contribute to the
non-parallelizable fraction and tipically uses system calls,
for details see~\cite{YavitsMulticoreAmdahl2014,CriticalSectionAmdahlEyerman:2010} and section~\ref{sec:Benchmarking}.


\section{The history of supercomputer development}\label{sec:SupercomputerHistory}

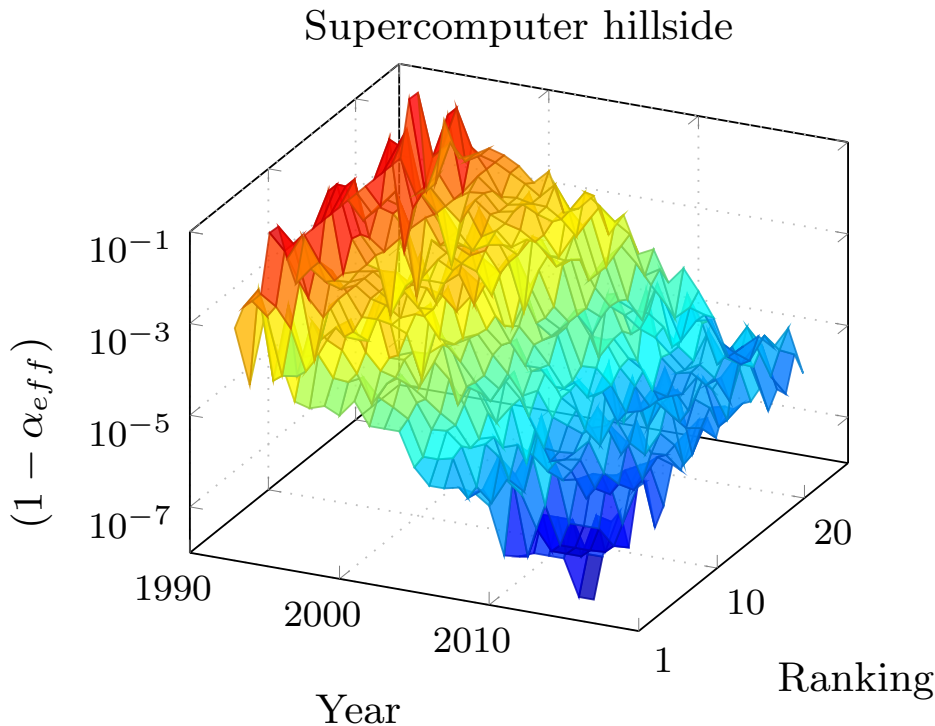
\begin{figure}
\begin{tikzpicture}[scale=1.75]
\pgfplotstableset{%
    col sep=semicolon,
    z index=0,
    y index=1,
    x index=2,
    header=false
}%
\pgfplotsset{
    every axis/.append style={
        scale only axis,
        width=\textwidth,
       height=\textwidth,
        xtick={1990,2000,2010},
        ytick={1,10,20},
        ztick={1e-7,1e-5,1e-3,1e-1}
    },
    /tikz/every picture/.append style={
        trim axis left,
        trim axis right,
    }
    }
   \begin{axis}
      [ 
    domain=1980:2020,
    domain y=1:40,
      footnotesize,
      title={Supercomputer hillside},
			/pgf/number format/.cd,
		use comma,
		1000 sep={},
		xlabel=Year,
		ylabel=Ranking,
		zlabel=$(1-\alpha_{eff})$,
		xmin=1990, xmax=2020,
		ymin=1, ymax=25, 
		zmin=1e-8, zmax=1e-1, 
		xlabel=Year,
		zmode = log,
		mesh/cols=25,
		grid=major,
		grid style={dotted},
		colormap/jet,
        zmode=log,
      ]

        \addplot3[%
            surf,
            opacity=0.8
        ] table {SupercomputerHillsideNew.csv};

\end{axis}
\end{tikzpicture}
\caption{The Top500 supercomputer parallelization efficiency.
The ($1-\alpha$) parameter for the past 25 years and the  (by $R_{max}$) first 25 computers. Data derived using the \gls{HPL} benchmark.
}
\label{SupercomputerHillside}

\end{figure}

As the supercomputer performance is traditionally characterized by the data
$R_{Max}$ and $R_{Peak}$, their efficiency can be derived
and using Equ.~(\ref{eq:alphafromr}) the value of $\alpha_{eff}$
can easily be calculated. 
Performing that calculation for the data in TOP500~\cite{TOP500:2017} for the first 25 supercomputers 
in the first 25 years, the "parallelization hillside"
for the top supercomputers can be derived, see Fig.~\ref{SupercomputerHillside}. From the figure
one can conclude that the value of $(1-\alpha_{eff})$
decreases with the number of years (as technology develops)
and increases with the ranking (the engineering ingenuity).

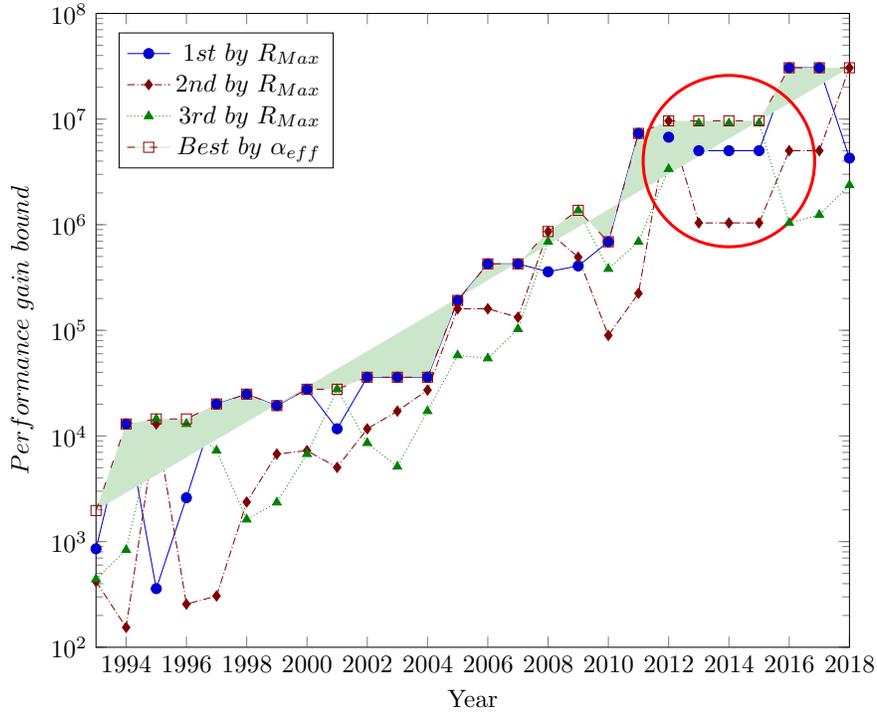
\begin{figure*}

\begin{tikzpicture}[scale=.95]
\begin{axis}
[
	width=\textwidth,
	cycle list name={my color list},
		legend style={
			cells={anchor=east},
			legend pos={north west},
		},
		xmin=1993, xmax=2018,
		ymin=1e2, ymax=1e8, 
		xlabel=Year,
		/pgf/number format/1000 sep={},
		ylabel=$Performance~gain~bound$,
		ymode=log,
		log basis x=2,
		]
\addplot table [x=a, y=b, col sep=comma] {Top500-0-Gain.csv};
		\addlegendentry{$1st~by~R_{Max}$}
\addplot table [x=a, y=c, col sep=comma] {Top500-0-Gain.csv};
		\addlegendentry{$2nd~by~R_{Max}$}
\addplot table [x=a, y=d, col sep=comma] {Top500-0-Gain.csv};
		\addlegendentry{$3rd~by~R_{Max}$}
\addplot table [x=a, y=e, col sep=comma] {Top500-0-Gain.csv};
		\addlegendentry{$Best~by~\alpha_{eff}$}

     \draw[very thick,red] (2014,4e6) circle(1.2cm);
\end{axis}
\end{tikzpicture}
\caption{The trend of the development
 of computing performance gain in the past 25 years, based on the 
first three (by $R_{Max}$) and the first (by ($1-\alpha$)) in the year in question. Data derived using the \gls{HPL} benchmark.
}
\label{SupercomputerTimelineGain}
\end{figure*}

Fig.~\ref{SupercomputerTimelineGain} displays the same data
from another point of view. On the vertical axis the performance gain (the single-processor performance is not included, see Equ.(\ref{eq:AmdahlMax})) is shown.
This diagram is for enhancing parallelization what
Moore observation is for enhancing electronic density,
i.e. \textit{the development of the parallelization of supercomputers is governed by the Amdahl's Law}.
Notice that the data are derived from the \textit{efficiency},
so neither single processor performance nor clock frequency are included.
It was theoretically concluded in section~\ref{sec:InherentLimit} that
the performance gain data are expected to saturate around the value
$10^7$, as displayed in the figure.

The data in the big circle show up a kind of stalling,
not experienced in the former years.
In other words, the saturation effect~\cite{ExponentialLawsComputing:2017} is displayed.
As was discussed in section~\ref{sec:InherentLimit},
the processor of $Taihulight$ deploys cooperative computing
(i.e. a slightly different computing principle) that keeps its achievable performance gain at a value slightly higher than the predicted maximum. 
The newly shined up supercomputer $Summit$, deploying  the conventional computing principle,
shows up the traditionally achievable parallelization gain; it could conquer slot \#1 only thanks to its better single processor performance.
The advantage of its processor for supercomputing was also underpinned by statistical considerations~\cite{Vegh:StatisticalConsiderations:2017}.

\section{A simplified model for parallel operation}\label{sec:ParallelModel}

\subsection{The performance losses}
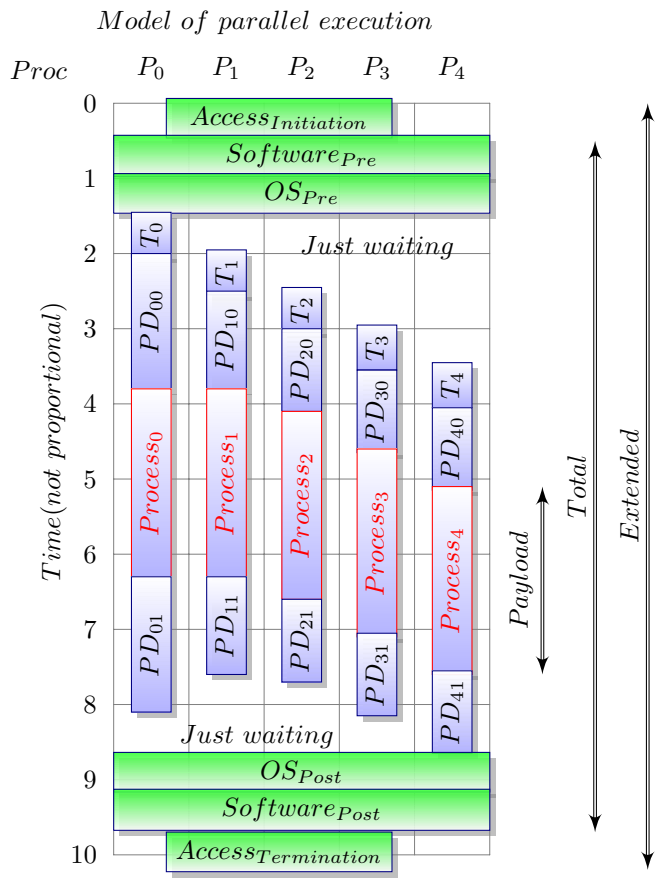
\begin{figure*}
	\begin{tikzpicture}[scale=1,cap=round]
	\node[right,above] at (11cm,.2cm) {$Proc$};
	\node[rotate=90,above=.5] at (11.5cm,-5cm) {$Time (not\  proportional)$};
	\foreach \y/\ytext in {0,...,10}
	\draw[yshift=-\y cm,xshift=-0.1cm+12cm]  node[left]
	{${\ytext}$};
	\draw[style=help lines,step=1] (12,-10) grid (17,0);
	\node[right,above] at (14cm,.8cm) {$Model\ of\ parallel\ execution$};
	\foreach \x/\xtext in {0,...,4}
	\draw[xshift=\x cm+12.5 cm,yshift=0.1cm]  node[above=.1]
	{$P_{\xtext}$};
	\QTAfigure{Access_{Initiation}}{.0}{12}{0}{0}{0}  %
	\QTCfigure{Software_{Pre}}{.0}{12}{0}{5}{0}  %
	\QTCfigure{OS_{Pre}}{.0}{12}{0}{10}{0}  %
	\QTfigure{T_0}{0.5}{12}{0}{15}{0}  %
	\QTfigure{PD_{00}}{1.8}{12}{0}{20}{0}  %
	\QTRfigure{Process_0}{2.5}{12}{0}{38}{0}  %
	\QTfigure{PD_{01}}{1.8}{12}{0}{63}{0}  %
	\QTfigure{T_1}{0.5}{13}{0}{20}{0}  %
	\QTfigure{PD_{10}}{1.3}{13}{0}{25}{0}  %
	\QTRfigure{Process_1}{2.5}{13}{0}{38}{0}  %
	\QTfigure{PD_{11}}{1.3}{13}{0}{63}{0}  %
	\QTfigure{T_2}{0.5}{14}{0}{25}{0}  %
	\QTfigure{PD_{20}}{1.1}{14}{0}{30}{0}  %
	\QTRfigure{Process_2}{2.5}{14}{0}{41}{0}  %
	\QTfigure{PD_{21}}{1.1}{14}{0}{66}{0}  %
	\QTfigure{T_3}{0.5}{15}{0}{30}{0}  %
	\QTfigure{PD_{30}}{1}{15}{0}{36}{0}  %
	\QTRfigure{Process_3}{2.5}{15}{0}{46}{0}  %
	\QTfigure{PD_{31}}{1}{15}{0}{71}{0}  %
	\QTfigure{T_4}{0.5}{16}{0}{35}{0}  %
	\QTfigure{PD_{40}}{1}{16}{0}{41}{0}  %
	\QTRfigure{Process_4}{2.5}{16}{0}{51}{0}  %
	\QTfigure{PD_{41}}{1}{16}{0}{76}{0}  %
	\node[right,above] at (15.5cm,-2.2cm) {$Just\ waiting$};
	\node[right,above] at (13.9cm,-8.7cm) {$Just\ waiting$};

	\QTCfigure{OS_{Post}}{.0}{12}{0}{87}{0}  %
	\QTCfigure{Software_{Post}}{.0}{12}{0}{92}{0}  %
	\QTAfigure{Access_{Termination}}{.0}{12}{0}{97.6}{0}  %
	
	%
	\coordinate (TP1) at (17.7 cm, -5.1);
	\coordinate (TP2) at (17.7 cm, -7.6);
	\draw[latex'-latex',double] (TP1) -- node[above, rotate=90] {$Payload$} (TP2);
	
	\coordinate (TT1) at (18.4 cm, -0.5);
	\coordinate (TT2) at (18.4 cm, -9.7);
	\draw[latex'-latex',double] (TT1) -- node[above,
	rotate=90] {$Total$} (TT2);
	
	\coordinate (TE1) at (19.1 cm, 0);
	\coordinate (TE2) at (19.1 cm, -10.2);
	\draw[latex'-latex',double] (TE1) -- node[above,
	rotate=90] {$Extended$} (TE2);
	\end{tikzpicture}
\caption{The extended Amdahl's model (strongly simplified)
}
\label{fig:Ourmodel}
\end{figure*}

 When speaking about computer performance,
a modern computer system 
is assumed, which comprises many sophisticated  components (in most cases embedding complete computers),
and their complex interplay results in the
final performance of the system. In the course of efforts to enhance processor performance through using some computing resources
in parallel, many ideas have been suggested and implemented, both in \gls{SW} and \gls{HW}~\cite{HwangParallelism:2016}.
All these approaches have different usage scenarios, performance and limitations. 
Because of the complexity of the task and the limited access to the components,
empirical methods and strictly controlled special measurement conditions are used to quantitize performance~\cite{LiljaComputerPerformance:2004}.
Whether a metric is appropriate for describing parallelism, depends on many factors~\cite{Karp:parallelperformance1990,
Sun:BetterPerformanceMetric1991,
ParallelEfficiency:Orii:2010}.

As mentioned in section~\ref{sec:reinterpret}, Amdahl listed different reasons why losses in the "computational load"
can occur. 
To understand operation of computing systems working in parallel, one needs to extend Amdahl's original (rather than that of the successors') model in such a way,
that non-parallelizable (i.e. apparently sequential) part comprises contributions from \gls{HW}, \gls{OS}, \gls{SW} and \gls{PD}\footnote{This separation cannot be strict.
Some features can be implemented in either \gls{SW} or \gls{HW}, or shared among them, and also some apparently sequential activities
may happen partly parallel with each other.}, and also some access time is needed for reaching the parallelized system.
The technical implementations of different parallelization methods show up infinite variety, so here a (by intention) strongly simplified model is presented. 
Amdahl's idea enables to put everything that cannot be parallelized into the sequential-only fraction.
The model is general enough to discuss qualitatively some examples of parallely working systems, neglecting different contributions
as possible in the different cases. The model can easily be converted to a technical (quantitative) one and the effect of inter-core communcation can also easily be considered. 

\subsection{The principle of the measurements}\label{sec:measprinciple}
When measuring performance, one faces serious difficulties, see for example~\cite{RISCVarchitecture:2017}, chapter 1,
both with making measurements and interpreting them. 
When making a measurement (i.e. running a benchmark) either on a single processor or a system of parallelized processors,
an instruction mix is executed many times. The large number of executions averages the rather different
execution times~\cite{Molnar:2017:Meas}, with an acceptable standard deviation.
In the case when the executed instruction mix is the same, the conditions (like cache and/or memory size,
the network bandwidth, \gls{I/O} operations, etc) are different and they form the subject of the comparison.
In the case when comparing different algorithms (like results of different benchmarks),
the instruction mix itself is also different.

Notice that the so called "algorithmic effects" -- like dealing with sparse data structures
(which affect cache behavior) or
communication between the parallelly running threads, like returning results repeatedly to the main thread
in an iteration (which greatly increases the non-parallelizable fraction in the main thread) -- 
 manifest through the \gls{HW}/\gls{SW} architecture,
and they can hardly be separated.
Also notice that there are fixed-size contributions, like utilizing time measurement facilities 
or calling system services. Since $\alpha_{eff}$ is a \textit{relative} merit, the \textit{absolute}
measurement time shall be long. When utilizing efficiency data from measurements which were
dedicated to some other goal, a proper caution must be exercised with the interpretation and accuracy
of the data.

\subsection{The formal introduction of the model}

The extended Amdahl's model is shown in Fig.~\ref{fig:Ourmodel}.
The contributions of the model component $XXX$ to $\alpha_{eff}$ 
will be denoted by $\alpha_{eff}^{XXX}$ in the followings.
Notice the different nature of those contributions.
They have only one common feature: \textit{they all consume time}.
The  vertical scale displays the actual activity for processing units shown on the horizontal scale. 

Notice that our model assumes no interaction between the processes
running on the parallelized systems in addition to the absolutely necessary minimum: starting and terminating the otherwise independent processes, which take parameters at the beginning and return results at the end.
It can, however, be trivially extended to the more general case when processes must share some resource (like a database, which shall provide different records for the different processes), 
either implicitly or explicitly. Concurrent objects have inherent sequentiality~\cite{InherentSequentiality:2012}, and synchronization and
communication among those objects 
considerably increase~\cite{YavitsMulticoreAmdahl2014} the non-parallelizable fraction
(i.e. contribution $(1-\alpha_{eff}^{SW})$), so \textit{in the case of extremely 
large number of processors special attention must be devoted to their role on the efficiency of the application} on the parallelized system.

Let us notice that all contributions have a role
during measurement: contributions due to \gls{SW}, \gls{HW}, ~\gls{OS} and  \gls{PD} cannot be separated,
though dedicated measurements can reveal their role, at least approximately.
 The relative weights of the different contributions are very different for the different parallelized systems,
and even within those cases depend on many specific factors,
so in every single parallelization case a careful analyzis is required.

\subsection{Access time}
Initiating and terminating the parallel processing is usually made from within the same computer,
except when one can only access the parallelized computer system from another computer (like in the case of clouds).
This latter access time is independent from the parallelized system,
and \textit{one must properly correct for the access time when derives timing
data for the parallelized system}. Amdahl's law is valid only for properly selected computing system.
This is a one-time, and usually fixed size time contribution.

\subsection{Execution time}

The execution time \textit{Total} covers all processings on the parallelized system.
All applications, running on a parallelized system, must make some non-parallelizable activity at least before beginning and after terminating parallelizable activity. 
This \gls{SW} activity represents what was assumed by Amdahl as the total sequential fraction\footnote{Although some \gls{OS} activity was surely included, Amdahl assumed some 20~\% \gls{SW} fraction, so the other contributions
could be neglected compared to \gls{SW} contribution.}.
As shown in Fig.~\ref{fig:Ourmodel}, the \textit{apparent} execution time includes the real payload activity, as well as waiting and \gls{OS}
and \gls{SW} activity. Recall that the execution times may be different~\cite{Molnar:2017:Meas,RISCVarchitecture:2017,hallaron} in the individual cases, even if the same processor executes the same instruction, but executing an instruction mix many times 
results in practically identical execution times, at least at model level. 
Note that the standard deviation of the execution times appears as a contribution to the non-parallelizable fraction,
and in this way increases the "imperfectness" of the architecture. This feature of \textit{processor}s deserves
serious consideration when utilizing a large number of processors.
Over-optimizing a processor for single-thread regime hits back when using it in a parallelized many-processor environment,
see also the statistical underpinning in~\cite{Vegh:StatisticalConsiderations:2017}.

\section{The non-parallelizable contributions}\label{sec:Contributions}

\subsection{The contribution of the operating system}
All applications must use \gls{OS} services and some \gls{HW} facilities to initialize themself as well as to access other processors.
Because operating system works in a different (supervisor) mode,
a considerable amount of time is required for the context switching\footnote{This is usually not a crucial contribution,
but under the extremal conditions represented by supercomputers, specialized operation systems must be used and every single core must run a lightweight \gls{OS}~\cite{CooperativeComputing2015}}.

The \gls{OS} initiates only the accessing of the processors, after that  \gls{HW} works partly in parallel with the next action of the \gls{OS}
and with the other actions initiating accessing other processors.
This period is denoted in Fig.~\ref{fig:Ourmodel} by $T_x$.
After the corresponding signals are generated, they must reach the target processor, that is they need some propagation time.
\gls{PD}s are denoted by $PD_{x0}$ and  $PD_{x1}$, corresponding to actions delivering input data and result, respectively.
This propagation time (which of course occurs in parallel with actions on other processors,
but which is a sequential contribution within the thread) depends strongly on how the processors are interconnected: 
this contribution can be considerable if the distance to travel is large or message transfer takes a long time
(like lengthy messages, signal latency, handshaking, store and forward operations in networks, etc.).

\subsection{The contribution of the physical size}

Although the signals travel in a computing system with nearly
the speed of the light, with increasing the physical size of the computer system a considerable time passes between issuing and receiving a signal,
causing the other party to wait, without making any payload job.
At the today's frequencies and chip sizes a signal cannnot even travel in one clock period from one side of the chip to the other, in the case of a stadium-sized supercomputer this delay can be in the order of several hundreds clock cycles.
Since the time of Amdahl, the ratio of the computing to the propagation time drastically changed, so --as~\cite{LimitsOfLimits2014} calls the attention to it--
it cannot be neglected any more, although presently it is not (yet) a dominating term.

\subsection{The contributions critical for large processor numbers}
Notice that the non-payload processing activity comprises two contributions of variable length, so handling them properly is crucial for high number of processors.
The first one is the \gls{OS} contribution loop iteration overhead:
processors must be handled one-by-one:
they receive arguments and return results.
This contribution simply increases linearly with the loop count. The second one is the propagation delay overhead \gls{PD} that increases with the physical size of the supercomputer.
The computer components can be in proximity in the range of $mm$, as well as in the range of $100~m$.
Similarly, they can be addressed in the first iteration or in the last one.
These two contributions may be very different for different processing units,
so combining the short ones from the first overhead class with long ones from the second overhead class reduces the overall overhead time, see Fig.~\ref{fig:Ourmodel}.

From the figure one can find out the meaning of the introduced metric $\alpha_{eff}$:
it is simply the ratio of the \textit{Payload} time
(the processors are utilized to do actual work) to the \textit{Total} execution time on the parallel system,
exactly as in the classic Amdahl model.
Notice that from the point of view of the processing units there is no difference \textit{why} they cannot do \textit{Payload} work:
all (in)activities are considered as contributions to the non-parallelizable fraction.
Notice also that using the \textit{Extended} time in place of the \textit{Total} time falsifies characteristics of the parameters of the parallelized system:
it adds some foreign contribution (time consumed by systems \textit{other} than the parallelized system) to the correct time.

\begin{figure*}
\def\constProcFreq{1}	
\def\constMPE{1} 
\def\constProcPerformance{(100*\constMPE)}  
\def\constNoOfProcessors{x*1e9/\constProcPerformance} 
\def\constTotalClocks{2e13} 
\def\constContextChange{1e4} 


\def\constMicroSecToTicks{1e3*\constProcFreq}  
\def\constAlphaContext{\constContextChange/\constTotalClocks}
\def\constAlphaLoop{\constNoOfProcessors/\constTotalClocks}
\def\constAlphaOS{\constAlphaContext
+\constAlphaLoop
}
\def\constAlphaTotal{(\constAlphaSW+\constAlphaOS)}
\def\constMinusAlpha{1-\constAlphaTotal}
\def\constEfficiency{(\constNoOfProcessors*\constAlphaTotal+\constMinusAlpha)}
\def\constRMax{x/\constEfficiency}
\def\constPropagationDelay{\constNoOfProcessors*1e-6/2e-8*2e9/\constTotalClocks*1000}



\maxsizebox{\textwidth}{!}
{{
\begin{tikzpicture}

\def\constAlphaSW{2e-8}

\pgfplotsset{
    xmin=0.001, xmax=1.1,
}

\begin{axis}[
  axis y line*=left,
  xlabel=$R_{Peak}(Eflop/s)$,
  ylabel=$(1-\alpha_{eff}^{HPL})$,
  ymin=1e-10, ymax=1e-5,
		xmode=log,
		log basis x=10,
		ymode=log,
		log basis y=10,
]
\addplot[samples=501,domain=.001:1.1,webbrown]
{\constAlphaSW }; \label{plot_SW}

\addplot[samples=501,domain=.001:1.1,webgreen]
{\constAlphaOS} ;\label{plot_loop}

%
\addplot[samples=501,domain=.001:1.1,webred,very thick]
{\constAlphaTotal} ;\label{plot_total}

\end{axis}

\begin{axis}[
  ylabel near ticks, yticklabel pos=right,
  axis x line=none,
    ymin=0.001, ymax=1.0,
		xmode=log,
		log basis x=10,
		ymode=log,
		log basis y=10,
  ylabel=$R_{Max}^{HPL}(Eflop/s)$,
		legend style={
			cells={anchor=east},
			legend pos={south east},
		},
]
\addlegendimage{/pgfplots/refstyle=plot_SW}
\addlegendentry{$\alpha^{SW}$}
\addlegendimage{/pgfplots/refstyle=plot_loop}
\addlegendentry{$\alpha^{OS}$}
\addlegendimage{/pgfplots/refstyle=plot_total}
\addlegendentry{$\alpha_{eff}$}
%

\addplot[samples=501,domain=.001:1.1,webblue,very thick] 
{\constRMax};\label{plot_rmax}

\addlegendimage{/pgfplots/refstyle=plot_rmax}
\addlegendentry{$R_{Max}(Eflop/s)$}

\end{axis}

\end{tikzpicture}
}
{
\begin{tikzpicture}

\def\constAlphaSW{2e-6}

\pgfplotsset{
    xmin=0.001, xmax=1.1,
}

\begin{axis}[
  axis y line*=left,
  xlabel=$R_{Peak}(Eflop/s)$,
  ylabel=$(1-\alpha_{eff}^{HPCG})$,
  ymin=1e-10, ymax=1e-5,
		xmode=log,
		log basis x=10,
		ymode=log,
		log basis y=10,
]
\addplot[samples=501,domain=.001:1.1,webbrown]
{\constAlphaSW }; \label{plot_SW}

\addplot[samples=501,domain=.001:1.1,webgreen]
{\constAlphaOS} ;\label{plot_loop}

%
\addplot[samples=501,domain=.001:1.1,webred,very thick]
{\constAlphaTotal} ;\label{plot_total}

\end{axis}

\begin{axis}[
  ylabel near ticks, yticklabel pos=right,
  axis x line=none,
    ymin=0.001, ymax=1.0,
		xmode=log,
		log basis x=10,
		ymode=log,
		log basis y=10,
  ylabel=$R_{Max}^{HPCG}(Eflop/s)$,
		legend style={
			cells={anchor=east},
			legend pos={south east},
		},
]
\addlegendimage{/pgfplots/refstyle=plot_SW}
\addlegendentry{$\alpha^{SW}$}
\addlegendimage{/pgfplots/refstyle=plot_loop}
\addlegendentry{$\alpha^{OS}$}
\addlegendimage{/pgfplots/refstyle=plot_total}
\addlegendentry{$\alpha_{eff}$}
%

\addplot[samples=501,domain=.001:1.1,webblue,very thick] 
{\constRMax};\label{plot_rmax}

\addlegendimage{/pgfplots/refstyle=plot_rmax}
\addlegendentry{$R_{Max}(Eflop/s)$}

\end{axis}

\end{tikzpicture}
}
}

%

%
%
\caption{Contributions $(1-\alpha_{eff}^X)$ to $(1-\alpha_{eff}^{total})$ and max payload performance $R_{Max}$ of a fictive supercomputer ( $P=1Gflop/s$ @ $1GHz$), imitating behavior of benchmarks HPL and HPCG.  The $(1-\alpha_{eff})$ values refer to the left scale, the $R_{Max}$ values to the right scale}
\label{fig:alphacontributions}
 \end{figure*}
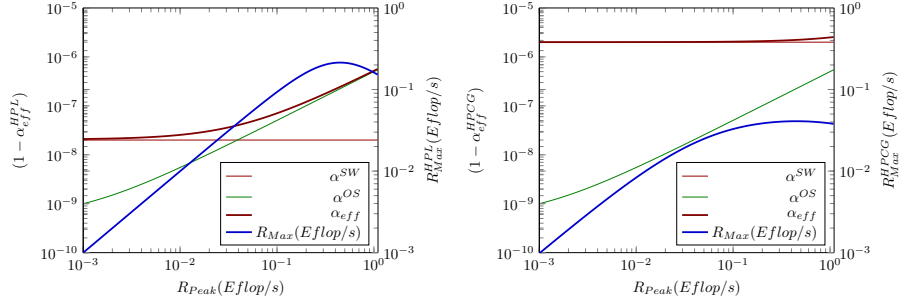
 
\subsection{How the dominance of contributions changes with the performance} 
Fig.~\ref{fig:alphacontributions} may help to understand the role of  both the different kinds of contributions $\alpha_{eff}^X$ and the benchmarks.
The figure displays diagrams for a hypotethic supercomputer\footnote{The technical model is not accurate enough because the needed parameters are not provided,
so only a guess for their order of magnitude can be given.},
with two parameter sets imitating the behavior of the benchmarks \gls{HPL} and \gls{HPCG}. For clarity, only the dominant contributions from \gls{OS} and \gls{SW} are shown.

The supercomputer is the same in both cases, the only essential difference is in the value of $\alpha^{SW}_{eff}$ (i.e. another benchmark runs on it, and utilizes the \gls{HW} facilities in a different way, see also section~\ref{sec:Benchmarking}).
As seen, the higher \gls{SW} contribution from the benchmark causes drastic changes
in the dependence of the payload performance on the nominal performance.
At low values of processor numbers (low payload performance) the deviation is hardly noticable. 
The higher \gls{SW} contribution decreases the achievable maximum $R_{Max}$,
and as the looping delay contribution due to the increasing number of processors exceeds the \gls{SW} contribution, the $R_{Max}$ diagrams decline.
In other words, as the nominal performance approaches the expected dream limit, the resulting $\alpha_{eff}$ starts to raise
and turns back the diagram of the actual computing performance $R_{Max}$.
This is a new phenomenon, noticable only at high number of processors.
At the time of Amdahl, $\alpha$ was considered as constant (and was dominated
by the \gls{SW} contribution), but in the exascale systems it tends to dominate.

The performance gain of supercomputers, see Fig.~\ref{SupercomputerTimelineGain},
shows up a behavior very similar to that of the Moore's law.
It looks like that  it increases year-by-year by a factor of cca. $1.5$ (the performance \textit{gain}, rather than the payload performance). 
Some reasons were presented, however, why also this behavior 
is not without limitations.
Let us recall, that inside the circle for three years there was no change in the parameters.
Even, with the appearance of the new world recorder in 2016 $Taihulight$ --which utilizes cooperative computing, a different principle-- the stall period seems to be prolonged for two more years.
Using some reasonable assumptions about the different contributions
to the non-parallelizable fraction mentioned above,
their order of magnitude were estimated, 
and also some saturation values were  predicted in section~\ref{sec:InherentLimit}.

Noticable that the performance gain in Fig.~\ref{fig:alphacontributions} even shows up a breakdown,
while  Fig.~\ref{SupercomputerTimelineGain} does not.
One can guess, however, that the stalling was caused by 
this breakdown: adding more processors \textit{decreases}
the actual performance, so no measurements data were published about the measurements with more processor and less performance; see also  section~\ref{sec:Perspectives}.

\section{Benchmarking supercomputer performance}\label{sec:Benchmarking}

As experienced in running the benchmarks \gls{HPL} and \gls{HPCG} and explained in connection with Fig.~\ref{fig:alphacontributions}, the different
benchmarks produce different payload performance 
and computational efficiency on the same supercomputer.
The model presented in Fig.~\ref{fig:Ourmodel} enables
to explain the difference.

The benchmarks, utilized to derive numerical parameters for supercomputers, are specialized \textit{program}s,
which run in the \gls{HW}/\gls{OS} \textit{environment} provided by the supercomputer under test.
One can use benchmarks for different goals. Two typical fields of utilization:
to describe  the environment supercomputer application runs in, and to guess
how quickly an application will run on a given supercomputer.

The (apparently) sequential fraction $(1-\alpha_{eff})$, as it is obvious from our model, cannot distinguish between
the (at least apparently) sequential processing time contributions of different origin,
even the \gls{SW} (including \gls{OS}) and \gls{HW} sequential contributions cannot be separated.
Similarly, it cannot be taken for sure that those contributions sum up linearly.
Different benchmarks provide different \gls{SW} contributions to
the non-parallelizable fraction of the execution time (resulting 
in different efficiencies and ranking~\cite{DifferentBenchmarks:2017}), 
so comparing results (and especially establishing ranking!)
derived using different benchmarks shall be done with maximum care.
\textit{Since the efficiency depends heavily on the number of cores, different configurations shall be compared using the same benchmark and the same number of processors (or same $R_{Peak}$).}

If the goal is to characterize the supercomputer's \gls{HW}+\gls{OS} system itself,
a benchmark program should distort \gls{HW}+\gls{OS} contribution  as little as possible, i.e. the
\gls{SW} contribution must be much lower than the \gls{HW}+\gls{OS} contribution.
In the case of supercomputers, the benchmark \gls{HPL} is used for this goal
since the beginning of the supercomputer age. 
The mathematical behavior of \gls{HPL} enables to 
minimize  \gls{SW} contribution, i.e.  \gls{HPL} \textit{delivers the
possible best estimation for $\alpha_{eff}^{HW+OS}$}.

If the goal is to estimate the expectable behavior of an application, 
the benchmark program should imitate the structure and behavior of the application. 
In the case of supercomputers, a couple of years ago the benchmark \gls{HPCG} 
has been introduced for this goal, since "\textit{\gls{HPCG} is designed to exercise computational and data access patterns that more closely match a different and broad set of important applications, and to give incentive to computer system designers to invest in capabilities that will have impact on the collective performance of these applications}"~\cite{HPCG_List:2016}. 
However, its utilization can be misleading: \textit{the ranking is only valid for the \gls{HPCG} application, and only utilizing that number of processors.}
\gls{HPCG} seems really to give better hints for designing supercomputer applications\footnote{This is why for example~\cite{DeepLearning2015} considers \gls{HPCG}  as "practical performance".}, than \gls{HPL} does.
According to our model, in the case of using the \gls{HPCG} benchmark, the \gls{SW} contribution dominates\footnote{
Returning calculated gradients requires much more 
sequential communication (unintended blocking).}, i.e.
 \gls{HPCG} delivers the
best estimation for $\alpha_{eff}^{SW}$ for this class of supercomputer applications.

Supercomputer community has extensively tested the efficiency of TOP500 supercomputers when benchmarked with \gls{HPL} and \gls{HPCG}~\cite{DongarraExascaleRace:2017}.
It was found that the efficiency (and $R_{Max}$) is typically 2 orders of magnitude lower when benchmarked with \gls{HPCG} rather than \gls{HPL}, even at relatively low number of processors.

\section{The efficiency of the applications}\label{sec:Efficiency}

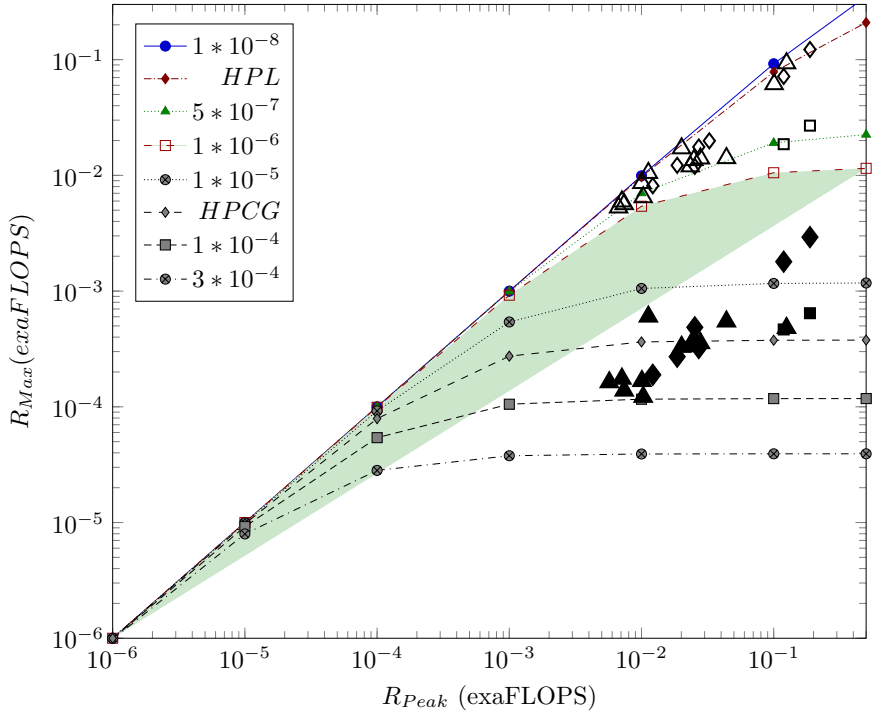
\begin{figure*}
\begin{tikzpicture}[scale=0.95]
\begin{axis}
[
	width=\textwidth,
	cycle list name={my color list},
		legend style={
			cells={anchor=east},
			legend pos={north west},
		},
		xmin=1e-6, xmax=0.5,
		ymin=1e-6, ymax=0.3, 
		xlabel={$R_{Peak}$  (exaFLOPS)},
		/pgf/number format/1000 sep={},
		ylabel={$R_{Max} (exaFLOPS)$},
		xmode=log,
		log basis x=10,
		ymode=log,
		log basis y=10,
		]
\addplot table [x=a, y=i, col sep=comma] {RMaxvsRPeakatAlpha.csv};
		\addlegendentry{$1*10^{-8}$}
\addplot table [x=a, y=h, col sep=comma] {RMaxvsRPeakatAlpha.csv};
		\addlegendentry{$HPL$}
\addplot table [x=a, y=k, col sep=comma] {RMaxvsRPeakatAlpha.csv};
		\addlegendentry{$5*10^{-7}$}
\addplot table [x=a, y=f, col sep=comma] {RMaxvsRPeakatAlpha.csv};
		\addlegendentry{$1*10^{-6}$}
\addplot table [x=a, y=e, col sep=comma] {RMaxvsRPeakatAlpha.csv};
		\addlegendentry{$1*10^{-5}$}
\addplot table [x=a, y=d, col sep=comma] {RMaxvsRPeakatAlpha.csv};
		\addlegendentry{$HPCG$}
\addplot table [x=a, y=c, col sep=comma] {RMaxvsRPeakatAlpha.csv};
		\addlegendentry{$1*10^{-4}$}
\addplot table [x=a, y=b, col sep=comma] {RMaxvsRPeakatAlpha.csv};
		\addlegendentry{$3*10^{-4}$}
 Supercomputers TOP10 by HPCG@2018 June
		\addplot[only marks,  mark=triangle,  mark size=4, thick] plot coordinates {
			(0.01127,0.0105) 
			(0.0439,0.0141) 
			(0.1007, 0.0614) 
			(0.125,0.0930) 
			(0.0249,  0.013554) 
			(0.0279,  0.0140) 
			(0.0234,  0.01197) 
			(0.0201,  0.01713) 
			(0.00711, 0.00595) 
			(0.01007, 0.00859) 
			(0.00671, 0.00528) 
			(0.00740, 0.00564) 
			(0.01030, 0.00647) 
		};
		\addplot[only marks,  mark=triangle*,  mark size=4, thick] plot coordinates {
			(0.01127,0.000602) 
 			(0.0439, 0.000546) 
			(0.125,  0.000480) 
			(0.0249, 0.000385) 
			(0.0279, 0.000355) 
			(0.0234, 0.000334) 
			(0.0201, 0.000330) 
			(0.00711,0.000175) 
			(0.01007,0.000167) 
			(0.00571,0.000163) 
			(0.00740,0.000138) 
			(0.01030,0.000122) 
			 	
		};
		\addplot[only marks,  mark=diamond,  mark size=3, thick] plot coordinates {
			(0.188,0.1223) 
			(0.119,0.0716) 
			(0.03258, 0.01988) 
			(0.0253,0.01196) 
			(0.02711, 0.01759) 
			(0.01862, 0.01221) 
			(0.01217, 0.00813) 
		};
		\addplot[only marks,  mark=square,  mark size=2, thick] plot coordinates {
			(0.188,0.0269) 
			(0.119,0.0186) 
		};
		\addplot[only marks,  mark=diamond*,  mark size=4, thick] plot coordinates {
			(0.188,  0.002926) 
			(0.119,  0.001795) 
			(0.0253, 0.000486) 
			(0.02711,0.000322) 
			(0.01862,0.000271) 
			(0.01217,0.000189) 
		};
		\addplot[only marks,  mark=square*,  mark size=2, thick] plot coordinates {
			(0.188,  0.000644) 
			(0.119,  0.000466) 
		};
\end{axis}
\end{tikzpicture}
\caption{The $R_{Max}$ payload performance in function of the peak performance $R_{Peak}$, at different $(1-\alpha_{eff})$ values. The figures 
display the measured values
derived using \gls{HPL} and \gls{HPCG} benchmarks, for the TOP15 supercomputers.
}
\label{fig:exaRMaxAlpha}
\end{figure*}

As discussed above, the value of $(1-\alpha_{eff})$ differs for the two 
famous benchmarks by more than two orders of magnitude.
For the users of exascale applications, it is primarily interesting
how their application  will run on an exascale supercomputer.
Fig.~\ref{fig:exaRMaxAlpha} attempts to orient them in this question. The shading shows how nonlinearly the
actual performance behaves when the nominal performance approaches 
the dream limit $1~Eflop/s$.

The figure displays how the payload performance depends
on the nominal performance, using $(1-\alpha_{eff})$ as parameter.
The diagram assumes that $\alpha_{eff}$ does not depend on the nominal performance,
and the nominal performance is set by changing virtually
the number of the cores.
For orientation, the best benchmark results (as of 2018 July) are shown for the supercomputers in the TOP10 (either by \gls{HPL} or by \gls{HPCG}).
The empty marks refer to the \gls{HPL} case, the filled ones to \gls{HPCG}. The diamonds denote GPU accelerated supercomputers, the triangles unaccelerated ones.

As predicted in section~\ref{sec:InherentLimit},
the \gls{HPL} payload performance data all fit in the $(1-\alpha_{eff})$ 
band $10^{-7}$\dots$10^{-6}$ (near to the $5*10^{-7}$ value)
and the \gls{HPCG}  payload peformance data all fit in the $\alpha_{eff}$ 
band $10^{-5}$\dots$10^{-4}$ (near to the $5*10^{-5}$ value).
The corresponding measured values in  Fig.~\ref{fig:exaRMaxAlpha} should be around the corresponding diagram lines marked by \gls{HPL} and \gls{HPCG}, repectively.
In the light of the model above, these experiences should be easy to comprehend.
The \gls{HPL} and \gls{HPCG} benchmarks measure the  $\alpha_{eff}^{HW}$ and $\alpha_{eff}^{SW}$, respectively.

The \gls{HPCG} case is easier to explain. 
The \gls{HPCG} measures the (dominant) contribution of the same software, so the measured performance value points are expected to scatter around the same value, provided that 
the single processor performance is not very different.
The full marks clearly show the saturation effect,
and the points scatter around the corresponding diagram line.
The exceptions are being the new champion and its small brother. The reason is the exceptional single-processor performance of their processors. If one corrects for the performance factor (with relative to the single-processor performance of the $Taihulight$), the agreement is perfect,
see the smaller filled quares at the corresponding nominal performance values.

In the \gls{HPL} case the marks seem to follow the predicted
\gls{HPL} diagram line. The top 4 supercomputers
 deploy the same (or quite similar) trick:
 they reduce the looping delay by organizing  the cores into groups.
Supercomputers $Summit$, $Sierra$ and $Tianhe-2$ organize single processors into clusters, $Taihulight$ organizes 
"clusters" inside the processor.
As  \gls{HPL} measures value of $\alpha_{eff}^{HW+OS}$, this benchmark is sensitive to decreasing the looping delay,
the dominant contribution.
This is why these supercomputers are on the top of the list.
In addition, the higher single-processor performance
also raises their value of $R_{Max}$.

The trick they use helps only, however, when the value of
$\alpha_{eff}^{HW+OS}$ represents a relatively  large contribution to the value of $\alpha_{eff}$.
In the \gls{HPL} case decreasing $\alpha_{eff}^{OS}$
by two-three orders of magnitude reduces considerably
the resulting effective parallelization,
so maybe up to an order of magnitude better value of $\alpha_{eff}$
can be measured.
In the case of \gls{HPCG}, however, the contribution of
$(1-\alpha_{eff}^{SW})$ is about two orders of magnitude 
larger than the contribution $(1-\alpha_{eff}^{HW+OS})$,
so decreasing this latter by orders of magnitude has only marginal effect on  $(1-\alpha_{eff})$. 
The computers deploying this "clusterization" trick
show up good results when benchmarked with \gls{HPL},
but not so good when measured with  \gls{HPCG}.
The exceptionally good values are made by that the performance is achieved through using much less processors
(due to utilizing accelerator).
See also Fig.~\ref{fig:alphacontributions}.

So it looks like that the payload performance of supercomputers 
is limited as predicted in section~\ref{sec:InherentLimit}.
Even for the  \gls{HPL}-class applications, only a few tenths of $Eflop/s$ can be achieved, for the more real-life
(\gls{HPCG}-class) applications achieving a few $Pflop/s$ 
can be a realistic target.

The major difference between those two classes is the $\alpha_{eff}^{SW}$ contribution,
mainly that the iterative nature of \gls{HPCG} requires intensive data exchange 
with the anchestor thread, repeating thread forking and joining
many times, as well as repeating calculation and communication of the new 
values for the next iteration many times.
\textit{All those actions increase the non-parallelizable fraction, i.e. decrease $\alpha_{eff}$ and $R_{Max}$.}
Programmers of exascale applications shall reorganize
their programs to mitigate this effect as much as possible:
a kind of "\gls{SW} clusterization" should be invented.
Making local data handling and calculations
in other than the achestor thread, as well as avoiding not strictly necessary communication should be advantageous.


\section{The perspectives of supercomputing}\label{sec:Perspectives}

\begin{figure*}
\begin{tikzpicture}[scale=.95]
\begin{axis}
[
	title={$R_{Max}$ of Top10 Supercomputers for benchmark \large $HPL$},
	width=\textwidth,
	cycle list name={my color list},
		legend style={
			cells={anchor=east},
			legend pos={north west},
		},
		xmin=0.005, xmax=1.1,
		ymin=0.003, ymax=0.3, 
		xlabel={$R_{Peak}$  (exaFLOPS)},
		/pgf/number format/1000 sep={},
		ylabel={$R_{Max} (exaFLOPS)$},
		xmode=log,
		log basis x=10,
		ymode=log,
		log basis y=10,
		]
\addplot table [x=a, y=b, col sep=comma] {RMaxHPL10.csv};
		\addlegendentry{Taihulight}
\addplot table [x=a, y=c, col sep=comma] {RMaxHPL10.csv};
		\addlegendentry{Tianhe-2}
\addplot table [x=a, y=d, col sep=comma] {RMaxHPL10.csv};
		\addlegendentry{Piz Daint}
\addplot table [x=a, y=e, col sep=comma] {RMaxHPL10.csv};
		\addlegendentry{Gyoukou}
\addplot table [x=a, y=f, col sep=comma] {RMaxHPL10.csv};
		\addlegendentry{Titan}
\addplot table [x=a, y=g, col sep=comma] {RMaxHPL10.csv};
		\addlegendentry{Sequoia}
\addplot table [x=a, y=h, col sep=comma] {RMaxHPL10.csv};
		\addlegendentry{Trinity}
\addplot table [x=a, y=i, col sep=comma] {RMaxHPL10.csv};
		\addlegendentry{Cori}
\addplot table [x=a, y=j, col sep=comma] {RMaxHPL10.csv};
		\addlegendentry{Oakforest}
\addplot table [x=a, y=k, col sep=comma] {RMaxHPL10.csv};
		\addlegendentry{K computer}
		\addplot[only marks,  mark=o,  mark size=3, thick] plot coordinates {
			(0.1254,0.09301) 
			(0.0549,0.033863) 
			(0.0253,0.01960) 
			(0.0282,0.01914) 
			(0.0271,0.01759) 
		    (0.0201,0.01711) 
		    (0.0439,0.01414) 
		    (0.0279,0.01401) 
		    (0.0249,0.01355) 
		    (0.0113,0.01051) 
		};
\end{axis}
\end{tikzpicture}
\caption{$R_{Max}$ performance of selected TOP10 (as of 2017 November) supercomputers
in function of their peak performance $R_{Peak}$, for the \gls{HPL} benchmark. The actual $R_{Max}$ values are denoted by a bubble.}
\label{fig:exaRMax}{}{}
\end{figure*}
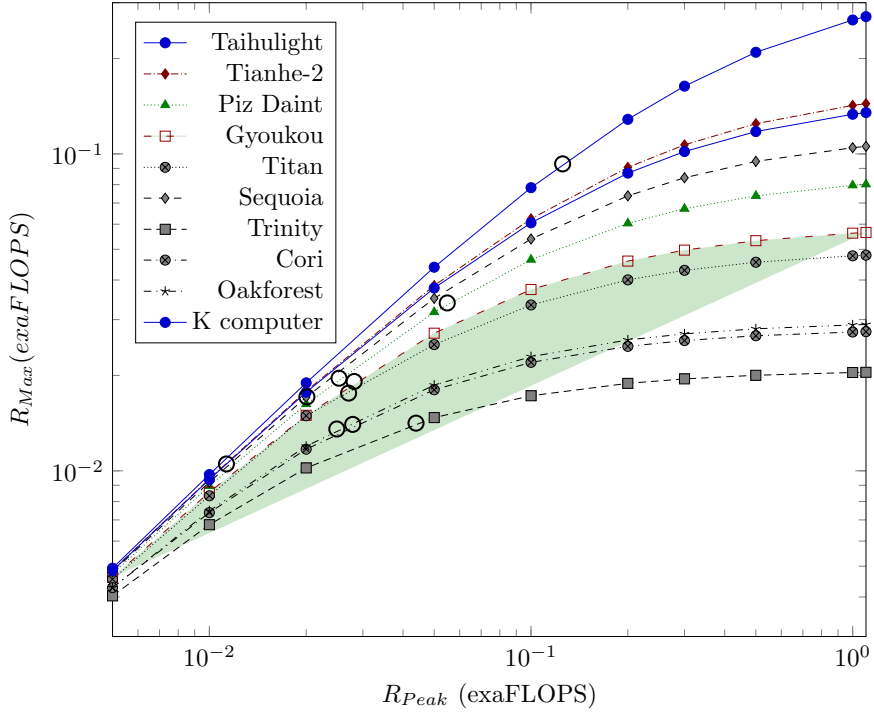

On the long term the "gold rush" will of course continue and
(partly due to the unusual strategic alliences)
new processors, connections, materials, principles will appear and maybe change the game.
Some short time predition, however, can be drawn from the tendencies and the
presently available data base, again changing virtually the number of processors in the TOP10 supercomputers, 
hoping that $\alpha_{eff}$ will not increase.
From this point of view, the predicted performance
values are optimistic: when no drastic changes happen, the shown future performance values will be surely not exceeded.
Fig.~\ref{fig:exaRMax} shows how the performance of the TOP10 supercomputers (as of November 2017)
would change with the conditions given above.
One can conclude that even the "benchmark payload performance" will not achieve the dream limit in the coming few years. See also Fig~\ref{fig:exaRMaxAlpha}.

One has to consider Fig.~\ref{fig:alphacontributions} seriously: simply increasing the number of the cores or utilizing accelerators is a short dead-end street. It is worth to re-read Amdahl's classic paper~\cite{AmdahlSingleProcessor67}: "\textit{the effort expended on achieving high parallel processing rates is wasted unless it is accompanied by achievements in sequential processing rates of very nearly the same magnitude}", see also Equ.~(\ref{eq:soverk}).
No single processor was added to the Chineese top supercomputers for years; $Gyoukou$ failed (and withdrawn) because it could only use 12\% of its processors;
$Aurora$ has been cancelled (and redesigned) because of similar reasons; $Summit$ utilizes only 2.3M cores out of the available 2.8M.
The newest (and probably worst) example is \textit{AI~Bridging~Cloud~Infrastructure} at slot \#5. They have outstanding single-processor performance but --because of using the cloud infrastructure-- a very poor parallelization ($\alpha_{eff}=5.9*10^{-5}$), so they can only use 2.8\% of their cores, although they have "only" 392K cores total.

\section{The role of the computing paradigm}\label{sec:Paradigm}

As it could be concluded from the discussion above,
one of the major issues in increasing the payload performance of supercomputers is that the looping delay should be decreased. 
This obstacle originates in the \gls{SPA}.
Since in the paradigm only one processor exists,
all components are designed in the spirit of \gls{SPA} and all supercomputers are build from commodity components.
As a consequence, on the bus only one processor can be addressed at a time and the cores cannot  directly exchange data between each other. 
The \gls{OS} takes over the responsibility of knowing about other processors, but it does it in a rather time-expensive way~\cite{Tsafrir:2007}.
The data exchange can take place only through some kind of "far" memory, causing slowdown ad sharing problems.

Reducing the looping delay can be relatively
easily solved by "clustering": one can organize the cores into "nodes" and then the main thread shall iterate only though the nodes rather than the cores.
In practice, it can be solved through organizing the single (many-core) processors into clusters, or delegating the cluster organization to the many-core processor~\cite{CooperativeComputing2015}.
Both solutions enabled a supercomputer to conquere the \#1 slot for a while. 

There are attempts to make direct data transfer through registers of different cores, like ~\cite{Congy:CoreSpilling:2007, ARM:big.LITTLE:2011}, but the idea of the explicit cooperation (in form of data transfer)
in processor chips has been implemented for the first time in~\cite{CooperativeComputing2015} (just notice that some 50 years later after that Amdahl suggested the idea~\cite{AmdahlSingleProcessor67}).
The power of the idea is clearly shown by that the supercomputer $Taihulight$ stayed on the top for two years,
when ranked by $R_{Max}^{HPL}$, and continues to stay when ranked be $\alpha_{eff}^{HPL}$ or $G^{HPL}$, thanks only to its processor.

Among others, the supercomputer experiences also underpin the need for renewing computing~\cite{RenewingComputingVegh:2018}.
Working out a computing paradigm which is drastically different from the traditional one and at the same time it is upward 
compatible with it, is not simple but possible~\cite{IntroducingEMPA2018}.
Its simulator~\cite{VeghEMPAthY86:2016} proves the feasibility and viability of the concept.
Utilizing that (or some similar) concept
the limitations stemming out from the computing paradigm can be circumvented, similarly to some other limitations of computing~\cite{LimitsOfLimits2014}.
In that way in the farther future the "dream limit" can be exceeded and exascale applications with reasonable efficiency can be prepared.

\section{Summary}\label{sec:Summary}

The paper discussed the exascale applications and exascale supercomputers as a complex \gls{HW}/\gls{SW} system.
It has pointed out that such systems have inherent performance limitations, and through understanding the reasons of those limitations, they can be mitigated.
The introduced naive model (based on extending Amdahl's principle) describes surprisingly well the performance values
measured on the recently announced supercomputers and explains some misteries experienced around supercomputers
approaching the exaFLOPS dream limit.
The model also explains the role of benchmarking and provides some practical hints for
writing efficient applications for the future exascale computers.
Based on the rigorously verified database of supercomputer performance data,
is is shown that with the present technologies (and computing principle)
the dream limit cannot be achieved. 
The need for "rebooting computing" is underlined by the analyzis and also a possible
way out of the present stalling through extending the computing paradigm is shown.

\section*{Acknowledgements}
Project no. 125547 has been implemented with the support provided from the National Research, Development and Innovation Fund of Hungary, financed under the K funding scheme.

\section*{References}


\end{document}